\def\isweb{1} 
\documentclass{pdg}

\reviewtitle{Cosmological Parameters}
\reviewauthor{O. Lahav (UCL) and A.R. Liddle (ULisboa)}
\reviewlabel{hubble}
\ischaptertrue

\global\mathchardef\LAMBDA="7003
\global\mathchardef\OMEGA="700A
\global\mathchardef\Delta="7101

\AtBeginDocument{%
  \renewcommand*{\Ref}[1]{%
     Ref.~\cite{#1}%
  }%
}

\usepackage{makeidx}
\makeindex

\ifdefined\isbooklet
\externalcitedocument{hubble-full}
\fi

\IfFileExists{crossref.aux}{\externaldocument{crossref}}{}

\begin{document}

% Set starting page number for book and booklet
\ifdefined\isbook
\setcounter{page}{467}
\fi
\ifdefined\isbooklet
\setcounter{page}{222}
\fi

% Enable bleeder tabs in book mode
\ifdefined\isbook
\rppthumb
\fi

% Include -main or -booklet file, depending on what we're making
\ifdefined\isbooklet
\fontsize{8pt}{9pt}\selectfont
\input{hubble-booklet}
\else
\begin{bibunit}
% Aug 2023 version (OL & AL)
% To latex: pdflatex hubble , bibtex hubble.1, pdflatex hubble,  pdflatex hubble

%!TEX root = hubble.tex

% $Id: BASENAME-main.tex 23686 2019-02-20 20:00:48Z beringer $
% Main file for PDG review hubble
%
% This file is included by the top-level file hubble.tex and is where the
% text of your review should be included. If desired, you may split your review into multiple
% files that are included from this file using \input.
%
% Do NOT modify the top-level file hubble.tex - it is generated from
% the PDG database and all manual changes WILL BE LOST!

% Review title
% ------------
% Please use \pdgtitle (rather than e.g. \chapter) to put the title of your review.
%
% To put the review title extracted from the PDG database, use \pdgtitle (no arguments).
% To override the default review title, you can use \pdgtitle[Some different title].
% In the latter case, please ask your overseer to update the review title in the database
\pdgtitle

% Table of contents
% -----------------
% If you want to include a table of contents at the start of your review,
% uncomment the line below. This should only be done for relatively long
% reviews.
%\tableofcontents

% Author information for this review
% ----------------------------------
% Please use one of the following forms:
%   \written{month year}
%   \revised{month year}
%   \customauthor{...}
% The first two, \written and \revised take the month and year when the review
% was written or revised as their only argument. The author list is generated
% from the PDG database. This is the preferred way of including author information.
% Only if really needed, you may use the third from, \customauthor, where you can
% specify the full text of the paragraph giving the review author information.
\revised{August 2023}

% Sectioning
% ----------
% This review is a regular review, please use \section for your top-level sectioning.
% For example:
%\section{Your first section title}

% Text of your review
% -------------------
% Please see the instructions at https://pdgdoc.lbl.gov/Pdg/ReviewTool on how to
% include figures, tables, and references. By default we include file examples.tex,
% which provides instructions and examples of how to include figures, tables, how
% to align equations, and more. It also produces an appendix showing all the standard
% PDG symbols and what they produce.
%
% To remove the examples from your review, comment out the following line when you
% start writing your review.
%\input{examples.tex}

%!TEX root = hubble.tex
%\BleederPointer=7
%% \font\tencaps=cmcsc10

%-----------------------------

% 2023 version, initiated 20-July-2023 by AL
%
%---------------------------
%\IndexEntry{hubblepage}

%% \chapter{Cosmological Parameters}

\index{Cosmology}%
\index{Distance--redshift relation}%
\index{Hubble!constant $H_0$}%
\index{Universe!Hubble expansion of}%

%Redefine upper case greek letters so that particle names are slanted
%Make new definitions to correspond to old definitions
%\global\mathchardef\LAMBDA="7003
%\global\mathchardef\OMEGA="700A
%\global\mathchardef\Delta="7101
%
\def\LAMBDA{{\rm \Lambda}}
\def\OMEGA{{\rm \Omega}}
%\beginRPPonly
%\runninghead{Cosmological parameters}
%\IndexEntry{hubbleCosmParam}
\index{Cosmological!parameters, note on}%

\def\paragraph#1{\medskip\noindent\bgroup\boldface\bfit #1:\egroup}
\def\mnras#1,#2(#3){{\rm MNRAS\ }{\bf #1}, {\rm#2} {\rm(#3)}}

\overfullrule=5pt

%\ifnum\WhichSection=7
%\magnification=\magstep1
%\fi

%Redefine upper case greek letters so that particle names are slanted
%Make new definitions to correspond to old definitions
%\WhoDidIt
%% {Updated August 2023, by O.\ Lahav (University College London) % 
%% and A.R.\ Liddle (Universidade de Lisboa).}%\WhoDidIt

%% \def\kmsMpc{km~s$^{-1}$~Mpc$^{-1}$}

%\ifnum\WhichSection=7
%\else
%\advance\baselineskip by -.2pt
%\fi

\def\ltsima{$\; \buildrel < \over \sim \;$}
\def\simlt{\lower.5ex\hbox{\ltsima}} \def\gtsima{$\; \buildrel > \over
\sim \;$} \def\simgt{\lower.5ex\hbox{\gtsima}}

\section{Parametrizing the Universe}
\label{hubble:sec}

\index{Parametrizing the Universe}%
\index{Universe!parametrizing}%
%\begin{figure}[h]
%\includegraphics[width=8cm]{H0_Verde.pdf}
%\end{figure}

Rapid advances in observational cosmology have led to the
establishment of a precision cosmological model, with many of the key
cosmological parameters determined to one or two significant figure
accuracy. Particularly prominent are measurements of cosmic microwave
background (CMB) anisotropies, with the highest precision observations
being those of the {\it Planck}
Satellite\cite{Planck:2018nkj,Planck:2018vyg}, which supersede the
landmark {\it WMAP}
results\cite{Bennett:2012zja,Hinshaw:2012aka}. However the most
accurate model of the Universe requires consideration of a range of
observations, with complementary probes providing consistency checks,
lifting parameter degeneracies, and enabling the strongest constraints
to be placed.

The term `cosmological parameters' now has a wide scope, and may
include the parameterization of some functions as well as simple
numbers describing properties of the Universe. The original usage
referred to the parameters describing the global dynamics of the
Universe, such as its expansion rate and curvature. Now we wish to
know how the matter budget of the Universe is built up from its
constituents: baryons, photons, neutrinos, dark matter, and dark
energy. We also need to describe the nature of perturbations in the
Universe, through global statistical descriptors such as the matter
and radiation power spectra. There may be additional parameters
describing the physical state of the Universe, such as the ionization
fraction as a function of time during the era since recombination.
Typical comparisons of cosmological models with observational data now
feature between five and ten parameters.

\subsection{The global description of the Universe}

\index{Universe!global description of}%
Ordinarily, the Universe is taken to be a perturbed 
\index{Robertson--Walker metric}%
Robertson--Walker space-time, with dynamics governed by Einstein's
equations. This is described in detail in the Big-Bang Cosmology
chapter in this volume.  Using the density parameters $\OMEGA_i$ for
the various matter species and $\OMEGA_\LAMBDA$ for the cosmological
constant, the Friedmann equation can be written
\begin{equation}\label{hubble.eq.redfried}
\sum_i \OMEGA_i + \OMEGA_\LAMBDA -1 = \frac{k}{R^2 H^2} \,,
\end{equation}
where the sum is over all the different species of material in the
Universe. This equation applies at any epoch, but later in this
article we will use the symbols $\OMEGA_i$ and $\OMEGA_\LAMBDA$ to
refer specifically to the present-epoch values.

The complete present-epoch state of the homogeneous Universe can be
described by giving the current-epoch values of all the density
parameters and the Hubble constant $h$ (the present-day Hubble
parameter being written $H_0 = 100 h \, {\rm km}\,{\rm s}^{-1}\,{\rm
Mpc}^{-1}$).  A typical collection would be baryons $\OMEGA_{{\rm
b}}$, photons $\OMEGA_\gamma$, neutrinos $\OMEGA_\nu$, and cold dark
matter $\OMEGA_{{\rm c}}$ (given charge neutrality, the electron
density is guaranteed to be too small to be worth considering
separately and is effectively included with the baryons).  The spatial
curvature can then be determined from the other parameters
using \Eq{hubble.eq.redfried}. The total present matter density
$\OMEGA_{{\rm m}} =\OMEGA_{{\rm c}}+\OMEGA_{{\rm b}}$ may be used in
place of the cold dark matter density $\OMEGA_{{\rm c}}$.

%\newpage % New page forced by hand as otherwise page 1 is over length
         % for some reason.

These parameters also allow us to track the history of the Universe,
at least back until an epoch where interactions allow interchanges
between the densities of the different species; this is believed to
have last happened at neutrino decoupling, shortly before Big-Bang
Nucleosynthesis (BBN).  To probe further back into the Universe's
history requires assumptions about particle interactions, and perhaps
about the nature of physical laws themselves.

The standard neutrino sector has three flavors. For neutrinos of mass
in the range $5 \times 10^{-4} \, {\rm eV}$ to $1\,{\rm MeV}$, the
density parameter in neutrinos is predicted to be
\begin{equation}
\OMEGA_\nu h^2 = \frac{\sum m_\nu}{93.12 \, {\rm eV}} \,,
\end{equation}
where the sum is over all families with mass in that range (higher
masses need a more sophisticated calculation). We use units with $c=1$
throughout. Results on atmospheric and Solar neutrino
oscillations\cite{Fukuda:2000np,*Ahmad:2001an} imply non-zero
mass-squared differences between the three neutrino flavors.  These
oscillation experiments cannot tell us the absolute neutrino masses,
but within the normal assumption of a mass hierarchy suggest a lower
limit of approximately $0.06 \, {\rm eV}$
%\IndexEntry{hubbleNuDens} 
\index{Neutrino(s)!mass density parameter, ${\rm \Omega}_{\nu}$}%
\index{Omeganu@${\rm \Omega}_{\nu}$, neutrino mass density parameter}%
for the sum of the neutrino masses (see the Neutrino chapter).

Even a mass this small has a potentially observable effect on the
formation of structure, as neutrino free-streaming damps the growth of
perturbations. Analyses commonly now either assume a neutrino mass sum
fixed at this lower limit, or allow the neutrino mass sum to be a
variable parameter. To date there is no decisive evidence of any
effects from either neutrino masses or an otherwise non-standard
neutrino sector, and observations impose quite stringent limits; see
the Neutrinos in Cosmology chapter.  However, we note that the
inclusion of the neutrino mass sum as a free parameter can affect the
derived values of other cosmological parameters.

\subsection{Inflation and perturbations}
%\IndexEntry{hubbleInflation}
\index{Inflation!of early Universe}%
\index{Universe!inflation}%

A complete model of the Universe must include a description of
deviations from homogeneity, at least statistically. Indeed, the most
powerful probes of the parameters described above come from the
evolution of perturbations, so their study is naturally intertwined
with the determination of cosmological parameters.

\index{Perturbation of early Universe}%
\index{Universe!perturbation}%

There are many different notations used to describe the perturbations,
both in terms of the quantity used and the definition of the
statistical measure. We use the dimensionless power spectrum
$\Delta^2$ as defined in the Big Bang Cosmology section (also denoted
${\cal P}$ in some of the literature). If the perturbations obey
Gaussian statistics, the power spectrum provides a complete
description of their properties.

From a theoretical perspective, a useful quantity to describe the
perturbations is the curvature perturbation ${\cal R}$, which measures
the spatial curvature of a comoving slicing of the space-time.  A
simple case is the
%\IndexEntry{hubbleHZeldovich}
\index{Harrison--Zeldovich spectrum}%
Harrison--Zeldovich spectrum, which corresponds to a constant
$\Delta^2_{{\cal R}}$. More generally, one can approximate the
spectrum by a power law, writing
\begin{equation}
\Delta^2_{{\cal R}}(k) = \Delta^2_{{\cal R}}(k_*) 
\left[\frac{k}{k_*}\right]^{n_{{\rm s}}-1} \,,
\end{equation}
where $n_{{\rm s}}$ is known as the spectral index, always defined so
that $n_{{\rm s}}=1$ for the Harrison--Zeldovich spectrum, and $k_*$
is an arbitrarily chosen scale.  The initial spectrum, defined at some
early epoch of the Universe's history, is usually taken to have a
simple form such as this power law, and we will see that observations
require $n_{{\rm s}}$ close to one. Subsequent evolution will modify
the spectrum from its initial form.

The simplest mechanism for generating the observed perturbations is
the inflationary cosmology, which posits a period of accelerated
expansion in the Universe's early
stages\cite{hubble:inf,hubble:PDP}. It is a useful working hypothesis
that this is the sole mechanism for generating perturbations, and it
may further be assumed to be the simplest class of inflationary model,
where the dynamics are equivalent to that of a single scalar field
$\phi$ with canonical kinetic energy slowly rolling on a potential
$V(\phi)$. One may seek to verify that this simple picture can match
observations and to determine the properties of $V(\phi)$ from the
observational data. Alternatively, more complicated models, perhaps
motivated by contemporary fundamental physics ideas, may be tested on
a model-by-model basis (see more in the Inflation chapter in this
volume).

Inflation generates perturbations through the amplification of quantum
fluctuations, which are stretched to astrophysical scales by the rapid
expansion. The simplest models generate two types, density
perturbations that come from fluctuations in the scalar field and its
corresponding scalar metric perturbation, and gravitational waves that
are tensor metric fluctuations. The former experience gravitational
instability and lead to structure formation, while both generate CMB
anisotropies.  Defining slow-roll parameters (with primes indicating
derivatives with respect to the scalar field, and $m_{\rm
Pl} \equiv \sqrt{\hbar c / G}$ the Planck mass) as
\begin{equation}
\epsilon = \frac{m_{{\rm Pl}}^2}{16\pi} \left( \frac{V'}{V} \right)^2 
\quad ,  \quad \eta = \frac{m_{{\rm Pl}}^2}{8\pi} \frac{V''}{V} \,,
\end{equation}
which should satisfy $\epsilon,|\eta| \ll 1$, the spectra can be
computed using the slow-roll approximation as
\begin{equation}
\Delta^2_{\cal R}(k)  \simeq  \left. \frac{8}{3 m_{{\rm Pl}}^4} \, 
\frac{V}{\epsilon} \right|_{k=aH} \quad , \quad
\Delta^2_{{\rm t}} (k) \simeq  \left. \frac{128}{3 m_{{\rm
Pl}}^4}  \, V \right|_{k=aH}\,.
\end{equation}
In each case, the expressions on the right-hand side are to be
evaluated when the scale $k$ is equal to the Hubble radius during
inflation. The symbol `$\simeq$' here indicates use of the slow-roll
approximation, which is expected to be accurate to a few percent or
better.

From these expressions, we can compute the spectral
indices\cite{Liddle:1992wi}:
\begin{equation}
n_{{\rm s}} \simeq 1-6\epsilon + 2\eta \quad ; \quad n_{{\rm t}} \simeq
-2\epsilon \,. 
\end{equation}
Another useful quantity is the ratio of the two spectra, defined by
\begin{equation}
r \equiv \frac{\Delta^2_{{\rm t}} (k_*)}{\Delta^2_{\cal R}(k_*)} \,.
\end{equation}
We have
\begin{equation}
r \simeq 16 \epsilon \simeq - 8 n_{{\rm t}} \,,
\end{equation}
which is known as the consistency equation.

One could consider corrections to the power-law approximation, which
we discuss later. However, for now we make the working assumption that
the spectra can be approximated by such power laws. The consistency
equation shows that $r$ and $n_{{\rm t}}$ are not independent
parameters, and so the simplest inflation models give initial
conditions described by three parameters, usually taken as
$\Delta^2_{{\cal R}}$, $n_{{\rm s}}$, and $r$, all to be evaluated at
some scale $k_*$, usually the `statistical center' of the range
explored by the data.  Alternatively, one could use the
parametrization $V$, $\epsilon$, and $\eta$, all evaluated at a point
on the putative inflationary potential.
 
After the perturbations are created in the early Universe, they
undergo a complex evolution up until the time they are observed in the
present Universe.  When the perturbations are small, this can be
accurately followed using a linear theory numerical code such as {\tt
CAMB} or {\tt CLASS}\cite{Lewis:1999bs,*Blas:2011rf}. This works right
up to the present for the CMB, but for density perturbations on small
scales non-linear evolution is important and can be addressed by a
variety of semi-analytical and numerical techniques. However the
analysis is made, the outcome of the evolution is in principle
determined by the cosmological model and by the parameters describing
the initial perturbations, and hence can be used to determine them.

Of particular interest are CMB anisotropies. Both the total intensity
and two independent polarization modes are predicted to have
anisotropies. These can be described by the radiation angular power
spectra $C_\ell$ as defined in the CMB article in this volume, and
again provide a complete description if the density perturbations are
Gaussian.

\subsection{The standard cosmological model}
%\IndexEntry{hubbleStandard}
\index{Standard cosmological model}%

\label{hubble.sub.params}

We now have most of the ingredients in place to describe the
cosmological model.  Beyond those of the previous subsections, we need
a measure of the ionization state of the Universe. The Universe is
known to be highly ionized at redshifts below 5 or so (otherwise
radiation from distant quasars would be heavily absorbed in the
ultra-violet), and the ionized electrons can scatter microwave
photons, altering the pattern of observed anisotropies. The most
convenient parameter to describe this is the optical depth to
scattering $\tau$ (\ie,~the probability that a given photon scatters
once); in the approximation of instantaneous and complete
reionization, this could equivalently be described by the redshift of
reionization $z_{{\rm i}}$.

As described in \Sec{hubble.sec.comb}, models based on these
parameters are able to give a good fit to the complete set of
high-quality data available at present, and indeed some simplification
is possible. Observations are consistent with spatial flatness, and
the inflation models so far described automatically generate
negligible spatial curvature, so we can set $k = 0$; the density
parameters then must sum to unity, and so one of them can be
eliminated. The neutrino energy density is often not taken as an
independent parameter; provided that the neutrino sector has the
standard interactions, the neutrino energy density, while
relativistic, can be related to the photon density using thermal
physics arguments, and a minimal assumption takes the neutrino mass
sum to be that of the lowest mass solution to the neutrino oscillation
constraints, namely $0.06 \, {\rm eV}$. In addition, there is no
observational evidence for the existence of tensor perturbations (with
the upper limits now starting to become constraining on models), and
so $r$ could be set to zero.  This leaves seven parameters, which is
the smallest set that can usefully be compared to the present
cosmological data. This model is referred to by various names,
including
%\IndexEntry{hubbleLCDM}
%\IndexEntry{hubbleConcord}
%\index{LambdaCDM@c$\Lambda$CDM, cold dark matter with dark energy}%
%\index{LambdaCDM@m${\rm \Lambda}$CDM, minimal cosmological model}%
\index{Concordance cosmology}%
$\LAMBDA$CDM, the 
concordance cosmology, and the standard cosmological model.

Of these parameters, only $\OMEGA_{\gamma}$ is accurately measured
directly.  The radiation density is dominated by the energy in the
CMB, and the 
\index{CMB!COBE satellite}%
\textit{COBE} satellite FIRAS experiment determined its temperature to be $T = 
(2.7255 \pm 0.0006) \, {\rm K}$\cite{Fixsen:2009ug},\footnote{All
quoted uncertainties in this article are $1\sigma$/68\% confidence and
all upper limits are 95\% confidence.  Cosmological parameters
sometimes have significantly non-Gaussian uncertainties. Values from
original sources have been rounded according to the conventions of
this volume.}  corresponding to $\OMEGA_{\gamma} = 2.47 \times 10^{-5}
h^{-2}$. It typically can be taken as fixed when fitting other
data. Hence the minimum number of cosmological parameters varied in
fits to data is six, though as described below there may additionally
be many `nuisance' parameters necessary to describe astrophysical
processes influencing the data.

In addition to this minimal set, there is a range of other parameters
that might prove important in future as the datasets further improve,
but for which there is so far no direct evidence, allowing them to be
set to specific values for now.  We discuss various speculative
options in the next section. For completeness at this point, we
mention one other interesting quantity, the helium fraction, which is
a non-zero parameter that can affect the CMB anisotropies at a subtle
level.  It is usually fixed in microwave anisotropy studies, but the
data are approaching a level where allowing its variation may become
mandatory.

In conventional parameter estimation, a set of parameters is chosen by
hand and the aim is to constrain their values. The higher-level
inference problem of model selection instead compares different
choices of parameter sets, as is necessary to assess whether
observations are pointing towards inclusion of new physical
effects. Bayesian inference offers an attractive framework for
cosmological model selection, setting a tension between model
predictiveness and ability to fit the data\cite{hubble:hobson}, and
its use is becoming widespread.

\subsection{Derived parameters}

The parameter list of the previous subsection is sufficient to give a
complete description of cosmological models that agree with
observational data. However, it is not a unique parameterization, and
one could instead use parameters derived from that basic
set. Parameters that can be obtained from the set given above include
the age of the Universe, the present horizon distance, the present
neutrino background temperature, the epoch of matter--radiation
equality, the epochs of recombination and decoupling, the epoch of
transition to an accelerating Universe, the baryon-to-photon ratio,
and the baryon-to-dark-matter density ratio.  In addition, the
physical densities of the matter components, $\OMEGA_i h^2$, are often
more useful than the density parameters.  The density perturbation
amplitude can be specified in many different ways other than the
large-scale primordial amplitude, for instance, in terms of its effect
on the CMB, or by specifying a short-scale quantity, a common choice
being the present linear-theory mass dispersion in a radius of $8 \,
h^{-1} {\rm Mpc}$, known as $\sigma_8$.

Different types of observation are sensitive to different subsets of
the full cosmological parameter set, and some are more naturally
interpreted in terms of some of the derived parameters of this
subsection than on the original base parameter set. In particular,
most types of observation feature degeneracies whereby they are unable
to separate the effects of simultaneously varying specific
combinations of several of the base parameters.

\section{Extensions to the standard model}
%\IndexEntry{hubbleExtensions}
\index{Extensions to the cosmological standard model}%

At present, there is no positive evidence in favor of extensions of
the standard model.  These are becoming increasingly constrained by
the data, though there always remains the possibility of trace effects
at a level below present observational capability.

\subsection{More general perturbations}

The standard cosmology assumes adiabatic, Gaussian
perturbations. Adiabaticity means that all types of material in the
Universe share a common perturbation, so that if the space-time is
foliated by constant-density hypersurfaces, then all fluids and fields
are homogeneous on those slices, with the perturbations completely
described by the variation of the spatial curvature of the
slices. Gaussianity means that the initial perturbations obey Gaussian
statistics, with the amplitudes of waves of different wavenumbers
being randomly drawn from a Gaussian distribution of width given by
the power spectrum. Note that gravitational instability generates
non-Gaussianity; in this context, Gaussianity refers to a property of
the initial perturbations, before they evolve.

The simplest inflation models, based on one dynamical field, predict
adiabatic perturbations and a level of non-Gaussianity that is too
small to be detected by any experiment so far conceived. For present
data, the primordial spectra are usually assumed to be power laws.

\subsubsection{Non-power-law spectra}

For typical inflation models, it is an approximation to take the
spectra as power laws, albeit usually a good one. As data quality
improves, one might expect this approximation to come under pressure,
requiring a more accurate description of the initial spectra,
particularly for the density perturbations. In general, one can expand
$\ln \Delta_{\cal R}^2$ as
\begin{equation}
\ln \Delta_{\cal R}^2(k) = \ln \Delta_{\cal R}^2(k_*) + (n_{{\rm s},*}-1) \ln 
\frac{k}{k_*} + \frac{1}{2} 
\left. \frac{dn_{{\rm s}}}{d\ln k} \right|_* \ln^2 \frac{k}{k_*} + \cdots \,,
\end{equation}
where the coefficients are all evaluated at some scale $k_*$. The term
$dn_{{\rm s}}/d\ln k|_*$ is often called the running of the spectral
index\cite{Kosowsky:1995aa}.  Once non-power-law spectra are allowed,
it is necessary to specify the scale $k_*$ at which the spectral index
is defined.

\subsubsection{Isocurvature perturbations}

An isocurvature perturbation is one that leaves the total density
unperturbed, while perturbing the relative amounts of different
materials. If the Universe contains $N$ fluids, there is one growing
adiabatic mode and $N-1$ growing isocurvature modes (for reviews
see \Ref{hubble:PDP} and \Ref{Malik:2008im}). These can be excited,
for example, in inflationary models where there are two or more fields
that acquire dynamically-important perturbations. If one field decays
to form normal matter, while the second survives to become the dark
matter, this will generate a cold dark matter isocurvature
perturbation.

In general, there are also correlations between the different modes,
and so the full set of perturbations is described by a matrix giving
the spectra and their correlations. Constraining such a general
construct is challenging, though constraints on individual modes are
beginning to become meaningful, with no evidence that any other than
the adiabatic mode must be non-zero.

\subsubsection{Seeded perturbations}

An alternative to laying down perturbations at very early epochs is
that they are seeded throughout cosmic history, for instance by
topological defects such as cosmic strings. It has long been excluded
that these are the sole original of structure, but they could
contribute part of the perturbation signal, current limits being just
a few percent\cite{Ade:2013xla}. In particular, cosmic defects
formed in a phase transition ending inflation is a plausible scenario
for such a contribution.

\subsubsection{Non-Gaussianity}

Multi-field inflation models can also generate primordial
non-Gaussianity (reviewed, \eg, in \Ref{hubble:PDP}). The extra fields
can either be in the same sector of the underlying theory as the
inflaton, or completely separate, an interesting example of the latter
being the curvaton
model\cite{Lyth:2001nq,*Enqvist:2001zp,*Moroi:2001ct}. Current upper
limits on non-Gaussianity are becoming stringent, but there remains
strong motivation to push down those limits and perhaps reveal trace
non-Gaussianity in the data. If non-Gaussianity is observed, its
nature may favor an inflationary origin, or a different one such as
topological defects.

\subsection{Dark matter properties}
%\IndexEntry{hubbleDMA}
\index{Dark matter}%
\index{OMEGA dm@${\rm \Omega}_{\rm dm}$, dark matter density}%

Dark matter properties are discussed in the Dark Matter chapter in
this volume.  The simplest assumption concerning the dark matter is
that it has no significant interactions with other matter, and that
its particles have a negligible velocity as far as structure formation
is concerned. Such dark matter is described as `cold,' and candidates
include the lightest supersymmetric particle, the axion, and
primordial black holes. As far as astrophysicists are concerned, a
complete specification of the relevant cold dark matter properties is
given by the density parameter $\OMEGA_{{\rm c}}$, though those
seeking to detect it directly need also to know its interaction
properties.

Cold dark matter is the standard assumption and gives an excellent fit
to observations, except possibly on the shortest scales where there
remains some controversy concerning the structure of dwarf galaxies
and possible substructure in galaxy halos.  It has long been excluded
for all the dark matter to have a large velocity dispersion, so-called
`hot' dark matter, as it does not permit galaxies to form; for thermal
relics the mass must be above about 1\,keV to satisfy this constraint,
though relics produced non-thermally, such as the axion, need not obey
this limit. However, in future further parameters might need to be
introduced to describe dark matter properties relevant to
astrophysical observations. Suggestions that have been made include a
modest velocity dispersion (warm dark matter) and dark matter
self-interactions. There remains the possibility that the dark matter
is comprised of two separate components, \eg, a cold one and a hot
one, an example being if massive neutrinos have a non-negligible
effect.

\subsection{Relativistic species}
%\IndexEntry{hubbleRelativisticSpecies}
\index{Hubble!relativistic species}%

The number of relativistic species in the young Universe (omitting
photons) is denoted $N_{\rm eff}$. In the standard cosmological model
only the three neutrino species contribute, and its baseline value is
assumed fixed at 3.044 (the small shift from 3 is because of a slight
predicted deviation from a thermal
distribution\cite{Bennett:2020zkv}). However other species could
contribute, for example an extra neutrino, possibly of sterile type,
or massless Goldstone bosons or other scalars. It is hence interesting
to study the effect of allowing this parameter to vary, and indeed
although 3.044 is consistent with the data, most analyses currently
suggest a somewhat higher value (\eg, \Ref{Riemer-Sorensen:2013iql}).

\subsection{Dark energy and modified gravity}
%\IndexEntry{hubbleDarkEnergy}
\index{Dark energy}%
\index{Modified gravity}%

While the standard cosmological model given above features a
cosmological constant, in order to explain observations indicating
that the Universe is presently accelerating, further possibilities
exist under the general headings of `dark energy' and `modified
gravity'. These topics are described in detail in the Dark Energy
chapter in this volume. This article focuses on the case of the
cosmological constant, since this simple model is a good match to
existing data. We note that more general treatments of dark
energy/modified gravity will lead to weaker constraints on other
parameters.

\subsection{Complex ionization history}

\index{Ionization history}%
\index{Ionization!history of the Universe}%
The full ionization history of the Universe is given by the ionization
fraction as a function of redshift $z$. The simplest scenario takes
the ionization to have the small residual value left after
recombination up to some redshift $z_{{\rm i}}$, at which point the
Universe instantaneously reionizes completely. Then there is a
one-to-one correspondence between $\tau$ and $z_{{\rm i}}$ (that
relation, however, also depending on other cosmological
parameters). An accurate treatment of this process will track separate
histories for hydrogen and helium.  While currently rapid ionization
appears to be a good approximation, as data improve a more complex
ionization history may need to be considered.
\index{Reionization of the Universe}%

\subsection{Varying `constants'}

\index{Hubble!varying constants}%
\index{Varying constants! gravitational $G_{{\rm N}}$, fine structure 
$\alpha$}% 
Variation of the fundamental constants of Nature over cosmological
times is another possible enhancement of the standard cosmology. There
is a long history of study of variation of the gravitational constant
$G_{{\rm N}}$, and more recently attention has been drawn to the
possibility of small fractional variations in the fine-structure
constant. There is presently no observational evidence for the former,
which is tightly constrained by a variety of measurements. Evidence
for the latter has been claimed from studies of spectral line shifts
in quasar spectra at redshift $z \simeq
2$\cite{Webb:2010hc,*King:2012id,**hubble:xxx}, but this is presently
controversial and in need of further observational study.

\subsection{Cosmic topology}

The usual hypothesis is that the Universe has the simplest topology
consistent with its geometry, for example that a flat Universe extends
forever.  Observations cannot tell us whether that is true, but they
can test the possibility of a non-trivial topology on scales up to
roughly the present Hubble scale. Extra parameters would be needed to
specify both the type and scale of the topology; for example, a
cuboidal topology would need specification of the three principal axis
lengths and orientation. At present, there is no evidence for
non-trivial cosmic topology\cite{Ade:2015bva}.
\index{Hubble!cosmic topology}%
\index{Cosmic topology}%

\section{Cosmological Probes}
\label{hubble.sec.obs}

The goal of the observational cosmologist is to utilize astronomical
information to derive cosmological parameters.  The transformation
from the observables to the parameters usually involves many
assumptions about the nature of the data, as well as of the dark
sector.  Below we outline the physical processes involved in each of
the major probes, and the main recent results. The first two
subsections concern probes of the homogeneous Universe, while the
remainder consider constraints from perturbations.

In addition to statistical uncertainties we note three sources of
systematic uncertainties that will apply to the cosmological
parameters of interest: (i) due to the assumptions on the cosmological
model and its priors (\ie,~the number of assumed cosmological
parameters and their allowed range); (ii) due to the uncertainty in
the astrophysics of the objects (\eg,~light-curve fitting for
supernovae or the mass--temperature relation of galaxy clusters); and
(iii) due to instrumental and observational limitations (\eg,~the
effect of `seeing' on weak gravitational lensing measurements, or beam
shape on CMB anisotropy measurements).

These systematics, the last two of which appear as `nuisance
parameters', pose a challenging problem to the statistical analysis.
We attempt a statistical fit to the whole Universe with 6 to 12
parameters, but we might need to include hundreds of nuisance
parameters, some of them highly correlated with the cosmological
parameters of interest (for example time-dependent galaxy biasing
could mimic the growth of mass fluctuations).  Fortunately, there is
some astrophysical prior knowledge on these effects, and a small
number of physically-motivated free parameters would ideally be
preferred in the cosmological parameter analysis.

\subsection{Measures of the Hubble constant}
\label{hubble:sec:MeasurHubconst}

In 1929, Edwin Hubble discovered the law of expansion of the Universe
by measuring distances to nearby galaxies.  The slope of the relation
between the distance and recession velocity is defined to be the
present-epoch Hubble constant, $H_0$.  Astronomers argued for decades
about the systematic uncertainties in various methods and derived
values over the wide range $40 \, {\rm km} \, {\rm s}^{-1} \, {\rm
Mpc}^{-1}\simlt H_0 \simlt 100 \, {\rm km} \, {\rm s}^{-1} \, {\rm
Mpc}^{-1}$.

%\IndexEntry{hubbleCepheid}
\index{HST, Hubble space telescope}%
\index{Cepheid variable stars} %
\index{Hubble!space telescope HST}%
One of the most reliable results on the Hubble constant came from the
Hubble Space Telescope \textit{(HST)} Key
Project\cite{Freedman:2000cf}.  This study used the empirical
period--luminosity relation for Cepheid variable stars, and calibrated
a number of secondary distance indicators:
%\IndexEntry{hubbleSupernovae}
%\index{Supernovae, Type Ia and Type II}%
Type~Ia Supernovae (SNe Ia); the Tully--Fisher relation;
surface-brightness fluctuations; and Type~II Supernovae.
\index{Type Ia supernovae}%
\index{Type II supernovae}%
\index{Supernovae!Type Ia}%
\index{Supernovae!Type II}%
Various systematics are under investigation, for example \textit{JWST}
is helping understanding of the effect of field crowding.  This
approach has been further extended, \eg using \textit{HST}
observations of Cepheids in the hosts of 42 SNe Ia and exploiting Gaia
EDR3 parallaxes, the SH0ES team derived $H_0 = (73.0 \pm1.0) \,{\rm
km} \, {\rm s}^{-1} \, {\rm Mpc}^{-1}$\cite{Riess:2021jrx}.

Three other methods have been used recently. One is a calibration of
 the tip of the red-giant branch applied to Type~Ia supernovae, the
 Carnegie--Chicago Hubble Programme (CCHP), finding $H_0 =
 (69.8 \pm 0.6 \, {\rm (stat.)} \pm 1.6 \, {\rm (sys.)}) \, {\rm km} \,
 {\rm s}^{-1} \, {\rm Mpc}^{-1}$\cite{Freedman:2021ahq}.  The second
 uses the method of time delay in gravitationally-lensed quasars;
 Birrer et al.~\cite{Birrer:2020tax} find $H_0 =
67.4^{+4.1}_{-3.2} \,{\rm km} \, {\rm s}^{-1} \, {\rm Mpc}^{-1}$ from
 a sample of 40 lenses. 
%(which includes as a subset the six systems previously used by the
%HOLiCOW collaboration\cite{Wong:2019kwg} who obtained a higher
%value).
A third method that came to fruition recently is based on
gravitational waves; the `bright standard siren' method applied to the
binary neutron star GW170817 yields $H_0 = 70^{+12}_{-8} \,{\rm km} \,
{\rm s}^{-1} \, {\rm Mpc}^{-1}$\cite{Abbott:2017xzu}.  Adding two
`dark standard siren' systems shifts this to $H_0 =
73^{+11}_{-8} \,{\rm km} \, {\rm s}^{-1} \, {\rm
Mpc}^{-1}$\cite{2023ApJ...943...56P}, still dominated by the single
bright siren system. The 47 sources in the third Gravitational-Wave
Transient Catalog (GWTC-3), again combined with the bright
siren, yield the similar result $H_0 = 68^{+12}_{-8} \,{\rm km} \,
{\rm s}^{-1} \, {\rm Mpc}^{-1}$\cite{LIGOScientific:2021aug}. When
many more gravitational-wave events have been acquired, the future
uncertainties on $H_0$ from standard sirens will get smaller.

\index{Collaborations!{\it Planck}}%
%\index{Planck collaboration@{\it Planck} collaboration}%
The determination of $H_0$ by the {\it Planck}
Collaboration\cite{Planck:2018vyg} gives a lower value than most of
the above methods, $H_0 = (67.4 \pm 0.5) \,{\rm km} \, {\rm s}^{-1} \,
{\rm Mpc}^{-1}$.  As they discuss, there is a strong degeneracy of $H_0$
with other parameters, particularly $\OMEGA_{\rm m}$ and the neutrino
mass.  It is worth noting that using the `inverse distance ladder'
method gives a result $H_0 = (67.8 \pm 1.3) \,{\rm km} \, {\rm
s}^{-1} \, {\rm Mpc}^{-1}$\cite{Macaulay:2018fxi}, close to the {\it
Planck} result.  The inverse distance ladder relies on
absolute-distance measurements from baryon acoustic oscillations
(BAOs) to calibrate the intrinsic magnitude of the SNe Ia (rather than
nearby Cepheids and parallax).  This measurement was derived from 207
spectroscopically-confirmed Type~Ia supernovae from the Dark Energy
Survey (DES), an additional 122 low-redshift SNe Ia, and measurements
of BAOs.  A combination of DES Year 3 (Y3) clustering and weak lensing
with BAO and BBN (assuming $\LAMBDA$CDM) gives $H_0 = (67.6 \pm
0.9) \,{\rm km} \, {\rm s}^{-1} \, {\rm Mpc}^{-1}$\cite{DES:2021wwk}.
The completed Extended Baryon Oscillation Spectroscopic Survey
(eBOSS)\cite{eBOSS:2020yzd}
%(REF: arXiv:2007.08991)
inverse distance ladder result, within an assumed extended
cosmological model, is $H_0 = (68.2 \pm 0.8) \,{\rm km} \, {\rm
s}^{-1} \, {\rm Mpc}^{-1}$, also close to the {\it Planck} value.

\index{Distance ladders}%
The tension between the $H_0$ values from {\it Planck} and the
traditional cosmic distance-ladder methods, with the SH0ES result
deviating from {\it Planck} by about $5\sigma$, is under intense
investigation for potential systematic effects.  There is possibly a
trend for higher $H_0$ derived from the nearby Universe and a lower
$H_0$ from the early Universe, which has led some researchers to
propose a time-variation of the dark energy component or other exotic
scenarios.  Ongoing studies are addressing the question of whether the
Hubble tension is due to systematics in at least one of the probes, or
a signature of new physics. Figure~\ref{hubble.fig.H0} shows a
selection of recent $H_0$ values, summarizing the current status of
the Hubble constant tension. See Ref.\cite{Shah:2021onj} for a review.

%\topfigure{H0}
%\pdgfigure{H0_plot_v2.pdf}%{center}{3in}%
\pdgfigure{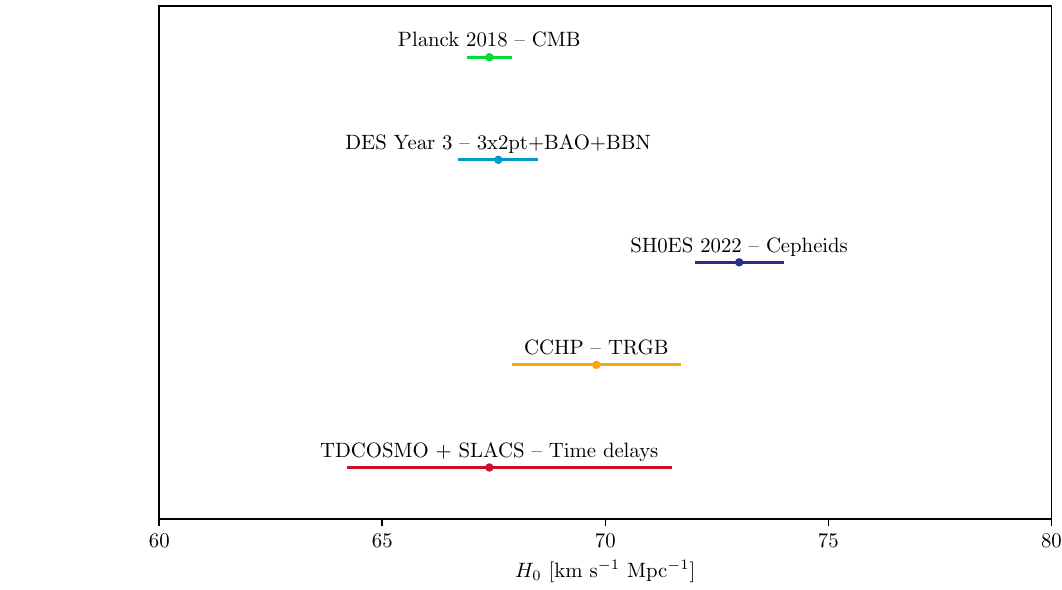}%{center}{3in}%
{A selection of recent $H_0$ measurements from the various projects as
described in the text. The standard-siren determinations are omitted
as they are too wide for the plot. Figure courtesy of Pablo Lemos.}
{hubble.fig.H0}{}{width=.90\columnwidth} ~\\ %To separate figure and
%caption better
%\IndexEntry{hubbleColorFigOne}
%\endfigure

%------------------------------------------------------------------------------------------------

\subsection{Supernovae as cosmological probes}

\index{Supernovae!as cosmological probes}%

\index{Cosmological probes!supernovae}%
Empirically, the peak luminosity of SNe Ia can be used as an efficient
distance indicator (\eg,~\Ref{Leibundgut:2001jd}), thus allowing
cosmology to be constrained via the 
\index{Distance--redshift relation}%
distance--redshift relation.  The
favorite theoretical explanation for SNe Ia is the thermonuclear
disruption of carbon--oxygen white dwarfs.  Although not perfect
`standard candles,' it has been demonstrated that by correcting for a
relation between the light-curve shape, color, and luminosity at
maximum brightness, the dispersion of the measured luminosities can be
greatly reduced.  There are several possible systematic effects that
may affect the accuracy of the use of SNe Ia as distance indicators,
\eg, evolution with redshift and interstellar extinction in
the host galaxy and in the Milky Way.

In the late 1990s two major studies, the Supernova Cosmology Project
and the High-$z$ Supernova Search Team, found evidence for an
\index{Accelerating Universe evidence}%
\index{Universe!accelerating evidence}%
accelerating
Universe\cite{Riess:1998cb,*Garnavich:1998th,*Perlmutter:1998np},
interpreted as due to a cosmological constant or a dark energy
component.  When combined with the CMB data (which indicate near
flatness, \ie, $\OMEGA_{{\rm m}} +
\OMEGA_{\LAMBDA} \simeq 1$), the best-fit values were $\OMEGA_{{\rm
m}} \simeq 0.3 $ and $\OMEGA_{\LAMBDA} \simeq 0.7 $.  Most results in
the literature are consistent with the $w=-1$ cosmological constant
case.  The leading sample currently is the Pantheon+
compilation\cite{Scolnic:2021amr}. This set of 1550
spectroscopically-confirmed SNe Ia gives $\OMEGA_{\rm m} = 0.334 \pm
0.018$ (stat+sym) for an assumed flat $\LAMBDA$CDM model, while in
combination with the CMB, for a flat $w$CDM model these data give
$w=-0.98^{+0.2}_{-0.3}$\cite{Brout:2022vxf}. 
%[REMOVED AS DES IS WITHIN PANTHEON+. ARL 20-Jul-23 ]For comparison,
%an analysis of a sample of 207 spectroscopically-confirmed DES SNe Ia
%combined with 122 low-redshift SNe\cite{Abbott:2018wog} yielded
%$\OMEGA_{\rm m} = 0.331 \pm 0.038$ for an assumed flat $\LAMBDA$CDM
%model.
Future experiments (\eg DES year 5 SNIa) will refine constraints on the
\index{Dark energy!equation of state parameter $w$}%
cosmic equation of state $w(z)$.

\subsection{Cosmic microwave background}
%\IndexEntry{hubbleCMB}
\index{CMB--Cosmic microwave background}%
\index{Cosmic microwave background, CMB}%

The physics of the CMB is described in detail in the CMB chapter in
this volume. Before recombination, the baryons and photons are tightly
coupled, and the perturbations oscillate in the potential wells
generated primarily by the dark matter perturbations. After
decoupling, the baryons are free to collapse into those potential
wells.  The CMB carries a record of conditions at the time of last
scattering, often called primary anisotropies. In addition, it is
affected by various processes as it propagates towards us, including
the effect of a time-varying gravitational potential (the integrated
Sachs-Wolfe effect), gravitational lensing, and scattering from
ionized gas at low redshift.

%\IndexEntry{hubbleCBranis}
%\IndexEntry{hubbleSW}
\index{Anisotropy of CMB}%
\index{CMB!anisotropy}%
\index{Sachs--Wolfe effect, integrated}%
\index{Integrated Sachs--Wolfe effect}%
The primary anisotropies, the integrated Sachs--Wolfe effect, and the
scattering from a homogeneous distribution of ionized gas, can all be
calculated using linear perturbation theory. Available codes include
{\tt CAMB} and {\tt CLASS}\cite{Lewis:1999bs}, the former widely used
embedded within the analysis package {\tt CosmoMC}\cite{Lewis:2002ah}
and in higher-level analysis packages such as {\tt
CosmoSIS}\cite{Zuntz:2014csq}, {\tt CosmoLike}\cite{Krause:2016jvl},
and Cobaya\cite{Torrado:2020dgo}.  Gravitational lensing is also
calculated in these codes. Secondary effects, such as inhomogeneities
in the reionization process, and scattering from
gravitationally-collapsed gas
%\IndexEntry{hubbleSZeldovich}
%\index{Zeldovich-Sunyaev effect}%
\index{Sunyaev--Zeldovich effect}%
(the Sunyaev--Zeldovich or SZ effect), 
require more complicated, and more uncertain, calculations.

The upshot is that the detailed pattern of anisotropies depends on all
of the cosmological parameters. In a typical cosmology, the anisotropy
power spectrum [usually plotted as $\ell(\ell+1)C_\ell/2\pi$] features a
flat plateau at large angular scales (small $\ell$), followed by a
series of oscillatory features at higher angular scales, the first and
most prominent being at around one degree ($\ell \simeq 200$). These
features, known as acoustic peaks, represent the oscillations of the
photon--baryon fluid around the time of decoupling. Some features can
be closely related to specific parameters---for instance, the location
in multipole space of the set of peaks probes the spatial geometry,
while the relative heights of the peaks probe the baryon density---but
many other parameters combine to determine the overall shape.

%\IndexEntry{hubblePlanck}\IndexEntry{hubbleWMAP}
\index{WMAP, Wilkinson Microwave Anisotropy Probe}%
The 2018 data release from the {\it Planck}
satellite\cite{Planck:2018nkj} gives the most powerful results to date
on the spectrum of CMB temperature anisotropies, with a precision
determination of the temperature power spectrum to beyond $\ell =
2000$. The Atacama Cosmology Telescope (ACT) and South Pole Telescope
(SPT) experiments extend these results to higher angular resolution,
though without full-sky coverage. {\it Planck} and the
polarization-sensitive versions of ACT and SPT give the current state
of the art in measuring the spectrum of $E$-polarization anisotropies
and the correlation spectrum between temperature and
polarization. These are consistent with models based on the parameters
we have described, and provide accurate determinations of many of
those parameters\cite{Planck:2018vyg}. Primordial $B$-mode
polarization has not been detected (although the gravitational lensing
effect on $B$ modes has been measured).

The data provide an exquisite measurement of the location of the set
of acoustic peaks, determining the angular-diameter distance of the
last-scattering surface. In combination with other data this strongly
constrains the spatial geometry, in a manner consistent with spatial
flatness and excluding significantly-curved Universes.  CMB data give
a precision measurement of the age of the Universe. The CMB also gives
a baryon density consistent with, and at higher precision than, that
coming from BBN. It affirms the need for both dark matter and dark
energy.  It shows no evidence for dynamics of the dark energy, being
consistent with a pure cosmological constant ($w = -1$).  The density
perturbations are consistent with a power-law primordial spectrum, and
there is no indication yet of tensor perturbations.
%\IndexEntry{hubbleReIon} 
\index{dA, angular-diameter distance@$d_A$, angular-diameter distance}%
\index{Reionization of the Universe}%
The current best-fit for the reionization optical depth from CMB data,
$\tau = 0.054$, is in line with models of how early structure
formation induces reionization.

{\it Planck} also made the first all-sky map of the CMB lensing field,
which probes the entire matter distribution in the Universe and adds
some additional constraining power to the CMB-only datasets. The
recent ACT results\cite{ACT:2023kun} demonstrate very well the
reconstruction of the mass power spectrum at intermediate redshifts
from CMB gravitational lensing.  These measurements are consistent
with {\it Planck}, supporting $\LAMBDA$CDM (see more in the CMB
chapter).

\subsection{Galaxy clustering}
%\IndexEntry{hubbleGalaxyClustering}
\index{Galaxy clustering}%
\label{hubble.sec.bias}

The power spectrum of density perturbations is affected by the nature
of the dark matter.  Within the $\LAMBDA$CDM model, the power spectrum
shape depends primarily on the primordial power spectrum and on the
combination $\OMEGA_{{\rm m}} h$, which determines the horizon scale
at matter--radiation equality, with a subdominant dependence on the
baryon density.  The matter distribution is most easily probed by
observing the galaxy distribution, but this must be done with care
since the galaxies do not perfectly trace the dark matter
distribution.  Rather, they are a `biased' tracer of the dark
matter\cite{Kaiser:1984sw}.  The need to allow for such bias is
emphasized by the observation that different types of galaxies show
bias with respect to each other.  In particular, scale-dependent and
stochastic biasing may introduce a systematic effect on the
determination of cosmological parameters from redshift
surveys\cite{Dekel:1998eq}.  Prior knowledge from simulations of
galaxy formation or from gravitational lensing data could help to
quantify biasing.  Furthermore, the observed 3D galaxy distribution is
in redshift space, \ie, the observed redshift is the sum of the Hubble
expansion and the line-of-sight peculiar velocity, leading to linear
and non-linear dynamical effects that also depend on the cosmological
parameters.  On the largest length scales, the galaxies are expected
to trace the location of the dark matter, except for a constant
multiplier $b$ to the power spectrum, known as the linear bias
parameter.  On scales smaller than 20 Mpc or so, the clustering
pattern is `squashed' in the radial direction due to coherent infall,
which depends approximately on the parameter
$\beta \equiv \OMEGA_{{\rm m}}^{0.6}/b$ (on these shorter scales, more
complicated forms of biasing are not excluded by the data).  On scales
of a few Mpc, there is an effect of elongation along the line of sight
(colloquially known as the `finger of God' effect) that depends on the
galaxy velocity dispersion.

\subsubsection{Baryon acoustic oscillations} 
%\IndexEntry{hubbleGalPwrSpec}
\index{Galaxy power spectrum}%
\index{Baryon!acoustic oscillations}%
\index{Baryon!oscillation spectroscopic survey}%
\index{BOSS, baryon oscillation spectroscopic survey}%

The power spectra of the 2-degree Field (2dF) Galaxy Redshift Survey
and the Sloan Digital Sky Survey (SDSS) are well fit by a $\LAMBDA$CDM
model and both surveys showed first evidence for baryon acoustic
oscillations (BAOs)\cite{Eisenstein:2005su,Cole:2005sx}.  When eBOSS
is combined with {\it Planck}, Pantheon Type~Ia Supernovae and other
probes the result is $w_{\rm p}=-1.020 \pm 0.032$ at the pivot redshift $z_{\rm p}
= 0.29$\cite{eBOSS:2020yzd}.  Similar results for $w$ were obtained
\eg by the WiggleZ survey\cite{Parkinson:2012vd}. Preliminary results
from the DESI galaxy redshift survey\cite{DESI:2023bgx} confirm the
detection of the BAOs, in agreement with previous surveys.

%The Baryon Oscillation Spectroscopic Survey (BOSS) of luminous red
%galaxies (LRGs) in the SDSS (DR 12) found, using a sample of 1.2
%million galaxies, consistency with $w=-1.01 \pm
%0.06$\cite{Alam:2016hwk} when combined with {\it Planck} 2015.
%Similar results for $w$ were obtained by the WiggleZ
%survey\cite{Parkinson:2012vd} [UPDATE TO eBOSS].

\subsubsection{Redshift distortion}
\index{Redshift distortion}%

There is continuing interest in Kaiser's `redshift distortion'
effect\cite{Kaiser:1987qv}.  This distortion depends on cosmological
parameters via the perturbation growth rate in linear theory $f(z) =
d \ln \delta /d \ln a \simeq \OMEGA_{\rm m}^\gamma(z)$, where $\gamma
\simeq 0.55$ for the $\LAMBDA$CDM model and may be different for 
modified gravity models.  By measuring $f(z)$ it is feasible to
constrain $\gamma$ and rule out certain modified gravity
models\cite{Guzzo:2008ac,Nusser:2011tu}. We note the
degeneracy of the redshift-distortion pattern and the geometric
distortion (the so-called Alcock--Paczynski
effect\cite{Alcock:1979mp}), %Fix by hand to remove space
                                %before full stop generated by \cite 
\eg, as illustrated by the WiggleZ
survey\cite{Blake:2012pj}
and eBOSS\cite{eBOSS:2020yzd}.
%and the BOSS Survey\cite{Gil-Marin:2016wya}. 

%\IndexEntry{hubbleSDSS}
\index{Sloan Digital Sky Survey (SDSS)}%

\subsubsection{Limits on neutrino mass from galaxy surveys and other
probes}
\label{hubble.sec.neut2dF}

%\IndexEntry{hubbleNuMass} 
\index{Neutrino(s)!mass, cosmological limit}%

Large-scale structure data place constraints on $\OMEGA_{\nu}$ due to
the neutrino free-streaming effect\cite{Lesgourgues:2006nd}.
Presently there is no clear detection, and upper limits on neutrino
mass are commonly estimated by comparing the observed galaxy power
spectrum with a four-component model of baryons, cold dark matter, a
cosmological constant, and massive neutrinos.  Such analyses also
assume that the primordial power spectrum is adiabatic,
scale-invariant, and Gaussian.  Potential systematic effects include
biasing of the galaxy distribution and non-linearities of the power
spectrum.  An upper limit can also be derived from CMB anisotropies
alone, while combination with additional cosmological datasets can
improve the results.

The most recent results on neutrino mass upper limits and other
neutrino properties are summarized in the Neutrinos in Cosmology
chapter in this volume.  The latest cosmological data constrain the
sum of neutrino masses to be below 0.13 eV\cite{DES:2021wwk}, consistent with
Refs.~\cite{Planck:2018vyg,eBOSS:2020yzd,ACT:2023kun}.  Since the lower limit on
this sum from oscillation experiments (assuming normal hierarchy) is
0.06 eV it is expected that future cosmological surveys will soon
detect effects from neutrino mass.
%While the latest cosmological data do not yet
%constrain the sum of neutrino masses to below 0.2 eV, since the lower
%limit on this sum from oscillation experiments is 0.06 eV it is
%expected that future cosmological surveys will soon detect effects
%from the neutrino mass.  
Also, current cosmological datasets are in good agreement with the
standard value for the effective number of neutrino species $N_{\rm
eff} = 3.044$.

\subsection{Clustering in the inter-galactic medium}
\index{Inter-galactic medium clustering}%

It is commonly assumed, based on hydrodynamic simulations, that the
neutral hydrogen in the inter-galactic medium (IGM) can be related to
the underlying mass distribution.  It is then possible to estimate the
matter power spectrum from the absorption observed in quasar spectra,
the so-called Lyman-$\alpha$ forest.  The usual procedure is to
measure the power spectrum of the transmitted flux, and then to infer
the mass power spectrum.  Photo-ionization heating by the ultraviolet
background radiation and adiabatic cooling by the expansion of the
Universe combine to give a simple power-law relation between the gas
temperature and the baryon density.  It also follows that there is a
power-law relation between the optical depth $\tau$ and $\rho_{{\rm
b}}$.  Therefore, the observed flux $F = \exp(-\tau)$ is strongly
correlated with $\rho_{{\rm b}}$, which itself traces the mass
density.  The matter and flux power spectra can be related by a
biasing function that is calibrated from simulations.  There are two
variants of Lyman-alpha analyses: 1-dimensional power spectra from
individual lines-of-sight that probe small ($\sim$Mpc) scales; and
3-dimensional Lyman-alpha BAO analyses that measure large-scale
correlations (over $\sim$100 Mpc scales) using neighboring quasar
lines-of-sight.

The latest BOSS and eBOSS datasets have provided measurements of the
BAO scale both in the Lyman-$\alpha$ absorption and in its
cross-correlation with quasars at an effective redshift
$z=2.3$\cite{duMasdesBourboux:2020pck,eBOSS:2020yzd}.  The results
agree within 1.5-$\sigma$ with the {\it Planck} CMB flat-$\LAMBDA$
estimation for BAO scale.  Such measurements will improve with the
\index{Dark energy spectroscopic instrument (DESI)}%
\index{DESI -- dark energy spectroscopic instrument}%
Dark Energy Spectroscopic Instrument (DESI) and other future
spectroscopic surveys.
%A study [ANY UPDATE?] of 266,590 quasars in the range $1.77 < z < 3$
% from SDSS was used to measure the BAO scale from the 3D correlation
% of Lyman-$\alpha$ and quasars\cite{Blomqvist:2019rah}.  Combined
% with the Lyman-$\alpha$ auto-correlation measurement presented in a
% companion paper\cite{Agathe:2019vsu} the BAO measurements at
% $z=2.34$ are within 1.7$\sigma$ of the {\it Planck} 2018
% $\LAMBDA$CDM model.
The Lyman-$\alpha$ flux power spectrum has also been used to constrain
the nature of dark matter, for example limiting the amount of warm
dark matter\cite{Viel:2013apy}.

\pdgfigure{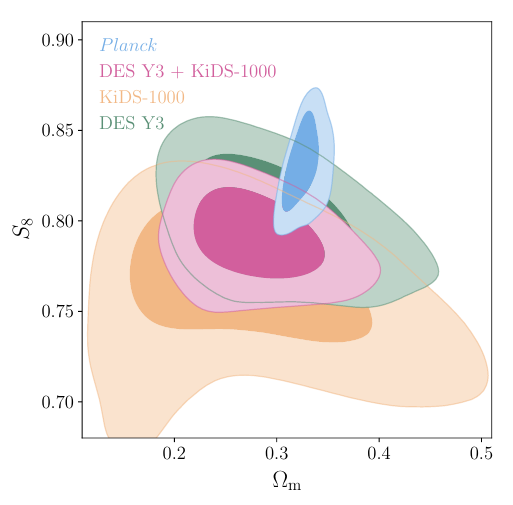}%{center}
{Marginalized posterior contours (inner 68\% confidence level, outer
 95\% confidence level) in the $\OMEGA_{{\rm m}}$--$S_8$ plane.  The
 plot shows KiDS weak lensing alone, DES Y3 weak lensing alone, DES
 and KiDS combined, and \textit{Planck} CMB. Figure courtesy of Alexandra Amon
 and the DES and KiDS collaborations.}
 {hubble.fig.DESY3_KiDS}{}{width=.75\columnwidth} ~\\ %To separate
 %figure and caption better

%\topfigure{KiDS} \pdgfigure{KiDS-S8Om.pdf}%{center} {Marginalized
%posterior contours (inner 68\% confidence level, outer 95\%
%confidence level) in the $\OMEGA_{{\rm m}}$--$S_8$ plane. Shown are
%the optical-only KiDS-450 analysis (green;
%{\protect \Ref{Hildebrandt:2018yau}}), the fiducial KiDS+VISTA-450
%setup (blue; {\protect \Ref{Hildebrandt:2018yau}}), DES Year 1 using
%cosmic shear only (purple; {\protect \Ref{Troxel:2017xyo}}), Troxel
%et al. 2018b, HSC-DR1 cosmic shear (orange;
%{\protect \Ref{hubble:Higake18}}) Hikage et al. 2018), and the {\it
%Planck} Legacy analysis (red; {\it Planck}
%Collaboration{\protect \cite{Planck:2018vyg}} using
%TT+TE+EE+lowE). Figure from {\protect \Ref{Hildebrandt:2018yau}}. }
%{hubble.fig.KiDS}{}{} ~\\ %To separate figure and caption
%better \IndexEntry{hubbleColorFigTwo} \endfigure

%+++++
\subsection{Weak gravitational lensing}
%\IndexEntry{hubbleGravLens}
\index{Gravitational lensing}%

Images of background galaxies are distorted by the gravitational
effect of mass variations along the line of sight.  Deep gravitational
potential wells, such as galaxy clusters, generate `strong lensing'
leading to arcs, arclets, and multiple images, while more moderate
perturbations give rise to `weak lensing.'  Weak lensing is now widely
used to measure the mass power spectrum in selected regions of the sky
(see \Ref{Refregier:2003ct,*Massey:2007wb,*Hoekstra:2008db} for
reviews).  Since the signal is weak, the image of deformed galaxy
shapes (the `shear map') must be analyzed statistically to measure the
power spectrum, higher moments, and cosmological parameters.  There
are various systematic effects in the interpretation of weak
lensing, \eg, due to atmospheric distortions during observations, the
redshift distribution of the background galaxies (usually depending on
the accuracy of photometric redshifts), the intrinsic correlation of
galaxy shapes, and non-linear modeling uncertainties.

Weak-lensing measurements from the Kilo-Degree Survey
(KiDS)\cite{KiDS:2020suj}, the Subaru Hyper-Suprime-Cam
(HSC)\cite{HSC:2018mrq}, and DES\cite{DES:2021vln} have constrained
the clumpiness parameter $S_8 \equiv \sigma_8 (\OMEGA_{{\rm
m}}/0.3)^{0.5}$.  Each of these surveys has yielded $S_8$ lower than
that derived from {\it Planck}. \Figure{hubble.fig.DESY3_KiDS} shows
the $\OMEGA_{{\rm m}}$--$S_8$ constraints derived from a joint
KiDS/DES analysis\cite{Kilo-DegreeSurvey:2023gfr} versus the CMB
constraint from {\it Planck}, where the tension is significantly
reduced, to $1.7\sigma$, as compared to previous analyses.  Results
from weak lensing from DES, combined with other probes, are described
in the next section.

%The `Kilo-Degree Survey' (KiDS), combined with the VISTA VIKING
%survey, used weak-lensing measurements over 450 deg$^2$ to constrain
%the clumpiness parameter $S_8 \equiv \sigma_8 (\OMEGA_{{\rm
%m}}/0.3)^{0.5} = 0.737^{+0.040}_{-0.036}$\cite{Hildebrandt:2018yau} .
%This is lower by $2.3\sigma$ than $S_8$ derived from {\it
%Planck}.  \Figure{hubble.fig.KiDS} (which is Figure 4
%from \Ref{Hildebrandt:2018yau}) shows the $\OMEGA_{{\rm m}}$--$S_8$
%constraints derived from weak lensing of KiDS, DES, and HPC versus
%the CMB constraint from {\it Planck}. Variations in $S_8$ among the
%weak-lensing surveys are mainly due to difference in the procedures
%for photometric-redshift determinations.  Results from weak lensing
%from DES, combined with other probes, are shown in the next section.

\subsection{Other probes}

Other probes that have been used to constrain cosmological parameters,
but that are not presently competitive in terms of accuracy, are the
\index{Cosmological probes!integrated Sachs-Wolfe effect}%
integrated Sachs-Wolfe effect\cite{Crittenden:1995xf,Planck:2015fcm},
the number density or composition of galaxy
clusters\cite{Ade:2015fva},
\index{Cosmological probes!galaxy peculiar velocities}%
and galaxy peculiar velocities that probe the mass fluctuations in
the local Universe\cite{Dekel:1994sx}.

\section{Bringing probes together}
\label{hubble.sec.comb}

% Maybe also say something about ?precision? vs ?accuracy? 

Although it contains two ingredients---dark matter and dark
energy---which have not yet been verified by laboratory experiments,
the $\LAMBDA$CDM model is almost universally accepted by cosmologists
as the best description of the present data. The approximate values of
some of the key parameters are $\OMEGA_{{\rm b}} \simeq 0.05$,
$\OMEGA_{{\rm c}} \simeq 0.25$, $\OMEGA_{\LAMBDA} \simeq 0.70$, and a
Hubble constant $h \simeq 0.70$. The spatial geometry is very close to
flat (and usually assumed to be precisely flat), and the initial
perturbations Gaussian, adiabatic, and nearly scale-invariant.
  
The most powerful data source is the CMB, which on its own supports
all these main tenets. Values for some parameters, as given
in \Ref{Planck:2018vyg}, are reproduced in
Table~\ref{hubble.tab.constraints}. These particular results presume a
flat Universe.  The constraints are somewhat strengthened by adding
additional datasets, BAO being shown in the Table as an example,
though most of the constraining power resides in the CMB data. Similar
constraints at lower precision were previously obtained by the {\it
WMAP} collaboration, and more recently by SPT.
\index{Collaborations!WMAP}%
%\index{WMAP collaboration}%

\def\tableheadsinglerule{\noalign{\medskip\hrule\smallskip}}
%\toptable{constraints}
\tabcolsep=0.11cm
\begin{pdgtable}{ccccccc}
{Parameter constraints reproduced from {\protect \Ref{Planck:2018vyg}}
(Table 2, column 5), with some additional rounding.  Both columns
assume the $\LAMBDA$CDM cosmology with a power-law initial spectrum,
no tensors, spatial flatness, a cosmological constant as dark energy,
and the sum of neutrino masses fixed to 0.06 eV. Above the line are
the six parameter combinations actually fit to the data ($\theta_{\rm
MC}$ is a measure of the sound horizon at last scattering); those
below the line are derived from these.  The first column uses {\it
Planck} primary CMB data plus the {\it Planck} measurement of CMB
lensing. This column gives our present recommended values. The second
column adds in data from a compilation of BAO measurements described
in {\protect \Ref{Planck:2018vyg}}.  The perturbation amplitude
$\Delta^2_{\cal R}$ (denoted $A_{\rm s}$ in the original paper) is
specified at the scale $0.05 \, {\rm Mpc}^{-1}$.  Uncertainties are
shown at 68\% confidence.}  {hubble.tab.constraints}{}
%\endCaption
%\centerline{\vbox{
%\tabskip=0pt\baselineskip 22pt
%\halign  {\strut#&\tabskip=1em plus 1em minus 0.5em% 
               %1st column is strut
%             \lefttab{#}&           % 2nd column is text FIRST
%             \centertab{#}&           % 2nd column is text FIRST
%             \centertab{#}&           % 3rd column is text FIRST
%             \centertab{#}\tabskip=0pt\cr % 7th text FOURTH
%\tableheaddoublerule%_______________________________________
%___________________________________________________________
\pdgtableheader{
&&   {\it \llap{ Pla}nck} TT,TE,EE+lowE+lens\rlap{ing}  & +BAO 
%\smallskip
}
%\tableheadsinglerule%_______________________________________
\cr
&$\OMEGA_{{\rm b}} h^2$ & $0.02237 \pm 0.00015$ & $0.02242 \pm 0.00014$ \cr
&$\OMEGA_{{\rm c}} h^2$ & $0.1200 \pm 0.0012$ & $0.1193 \pm 0.0009$ \cr
&$100 \,\theta_{\rm MC}$ & $1.0409 \pm 0.0003$ & $1.0410 \pm 0.0003$ \cr
&$n_{{\rm s}}$ & $0.965 \pm 0.004$ & $0.966 \pm 0.004$ \cr
&$\tau$ & $0.054 \pm 0.007$ & $0.056 \pm 0.007$ \cr
&$\ln(10^{10}\Delta^2_{\cal R})$ & $3.044 \pm 0.014$ & $3.047 \pm 0.014$ \cr
\tableheadsinglerule

&$h$ & $0.674 \pm 0.005$ & $0.677\pm 0.004$ \cr
&$\sigma_8$ & $0.811 \pm 0.006$ & $0.810 \pm 0.006$ \cr
&$\OMEGA_{{\rm m}}$ & $0.315 \pm 0.007$ & $0.311 \pm 0.006$ \cr
&$\OMEGA_\LAMBDA$ & $0.685 \pm 0.007$ & $0.689 \pm 0.006$  \cr
%\tablefootdoublerule%_______________________________________
%___________________________________________________________
\end{pdgtable}

If the assumption of spatial flatness is lifted, it turns out that the
primary CMB on its own constrains the spatial curvature fairly weakly,
due to a parameter degeneracy in the angular-diameter
distance. However, inclusion of other data readily removes this
degeneracy. Simply adding the {\it Planck} lensing measurement, and
with the assumption that the dark energy is a cosmological constant,
yields a 68\% confidence constraint on
%\IndexEntry{hubbleTotDen}
\index{OMEGAtot@${\rm \Omega}_{\rm tot}$, total energy density of Universe}%
\index{Total energy density of Universe, ${\rm \Omega}_{\rm tot}$}%
 $\OMEGA_{{\rm tot}} \equiv 
\sum \OMEGA_i + \OMEGA_\LAMBDA  = 1.011 \pm
0.006$ and further adding BAO makes it $0.9993 \pm
0.0019$\cite{Planck:2018vyg}.  Results of this type are normally taken
as justifying the restriction to flat cosmologies.

%\IndexEntry{hubbleAgeUniverse}
\index{Universe!age of}%
\index{Flatness of Universe}%
\index{Universe!flatness}%

One derived parameter that is very robust is the age of the Universe,
since there is a useful coincidence that for a flat Universe the
position of the first peak is strongly correlated with the age.  The
CMB data give $13.797 \pm 0.023$ Gyr (assuming flatness).  This is in
good agreement with the ages of the oldest globular clusters and with
radioactive dating.
 
The baryon density $\OMEGA_{\rm b}$ is now measured with high accuracy
from CMB data alone, and is consistent with and more precise than the
determination from BBN.  The value quoted in the Big-Bang
Nucleosynthesis chapter in this volume is $\OMEGA_{{\rm b}} h^2 =
0.0220 \pm 0.0004$.

While $\OMEGA_\LAMBDA$ is measured to be non-zero with very high
confidence, there is no evidence of evolution of the dark energy
density.  As described in the Dark Energy chapter in this volume a
combination of CMB, SN, RSD, and BAO measurements, assuming a flat
Universe, yield $w=-1.03 \pm 0.03$\cite{DES:2021wwk}, consistent with
the cosmological constant case $w = -1$.  Allowing more complicated
forms of dark energy weakens the limits.
  
The data provide strong support for the main predictions of the
simplest inflation models: spatial flatness and adiabatic, Gaussian,
nearly scale-invariant density perturbations. But it is disappointing
that there is no sign of primordial gravitational waves; combining
{\it Planck} and {\it WMAP} with BICEP2/Keck Array BK18 data (plus BAO
data to help constrain $n_{\rm s}$) gives a 95\% confidence upper
limit of $r<0.036$ at the scale $0.05 \, {\rm
Mpc}^{-1}$\cite{BICEPKeck:2021gln}. The density perturbation spectral
index is clearly required to be less than one by current data, though
the strength of that conclusion can weaken if additional parameters
are included in the model fits.

Tests have been made for various types of non-Gaussianity, a
particular example being a parameter $f_{{\rm NL}}$ that measures a
quadratic contribution to the perturbations. Various non-Gaussian
shapes are possible (see \Ref{Planck:2019kim} for details), and
current constraints on the popular `local', `equilateral', and
`orthogonal' types (combining CMB temperature and polarization data)
are $ f^{{\rm local}}_{{\rm NL}} = -1 \pm 5$, $f^{{\rm equil}}_{{\rm
NL}} = -26 \pm 47$, and $f^{{\rm ortho}}_{{\rm NL}} = -38 \pm 24$,
respectively (these may appear weak, but prominent non-Gaussianity
requires the product $f_{{\rm NL}}\Delta_{\cal R}$ to be large, and
$\Delta_{\cal R}$ is of order $10^{-5}$). Clearly none of these give
any indication of primordial non-Gaussianity.

While the above results come from the CMB alone, other probes are
becoming competitive (especially when considering more complex
cosmological models), and so the combination of data from different
sources is of growing importance.  We note that it has become
fashionable to combine probes at the level of power-spectrum data
vectors, taking into account nuisance parameters in each type of
measurement.  Discussions on `tension' in resulting cosmological
parameters depend on the statistical approaches used. Commonly the
cosmology community works within the Bayesian framework, and assesses
agreement amongst data sets with respect to a model via Bayesian
evidence, essentially the denominator in Bayes' theorem.

%Re-arranged text (OL, 20/8/2021) Recent examples include
%KiDS+GAMA \cite{vanUitert:2017ieu} and \index{Dark energy!survey
%(DES)}% Dark Energy Survey (DES) \cite{DES:2021wwk}. For example, the
%DES analysis includes galaxy position--position clustering,
%galaxy--galaxy lensing, and weak lensing shear.

As an example of results, combining DES Y3 (position--position
clustering, galaxy--galaxy lensing, and weak lensing shear) with {\it
Planck}, BAO, and RSD measurements from eBOSS, and Type~Ia supernovae
from the
%\index{Joint lightcurve analysis (JLA)}%
Pantheon dataset compilation, has shown the datasets to be mutually
compatible and yields very tight constraints on cosmological
parameters: $S_8 \equiv \sigma_8 (\OMEGA_{{\rm m}}/0.3)^{0.5} =
0.812 \pm 0.008$ and $\OMEGA_{{\rm m}} = 0.306^{+0.004}_{-0.005}$ in
$\LAMBDA$CDM, and $w=-1.03 \pm 0.03$ in $w$CDM\cite{DES:2021wwk}
matching the constraint in Ref.~\cite{eBOSS:2020yzd}. The combined
measurement of the Hubble constant within $\LAMBDA$CDM gives $H_0=
68.0^{+0.4}_{-0.3} \;{\rm km} \, {\rm s}^{-1} \, {\rm Mpc}^{-1}$,
still leaving substantial tension with the SH0ES measurement described
earlier. Future analyses and the next generation of surveys will test
for deviations from $\LAMBDA$CDM, for example epoch-dependent $w(z)$
and modifications to General Relativity.

\section{Outlook for the future}

\index{Cosmology!outlook for the future}%

The concordance model is now well-established, and there seems little
room left for any dramatic revision of this paradigm. A measure of the
strength of that statement is how difficult it has proven to formulate
convincing alternatives.

Should there indeed be no major revision of the current paradigm, we
can expect future developments to take one of two directions. Either
the existing parameter set will continue to prove sufficient to
explain the data, with the parameters subject to ever-tightening
constraints, or it will become necessary to deploy new parameters. The
latter outcome would be very much the more interesting, offering a
route towards understanding new physical processes relevant to the
cosmological evolution. There are many possibilities on offer for
striking discoveries, for example:
\begin{itemize}
\item the cosmological effects of a neutrino mass may be unambiguously
detected, shedding light on fundamental neutrino properties;
\item detection of primordial non-Gaussianities would indicate that
non-linear processes influence the perturbation generation mechanism;
\item detection of variation in the dark energy density (\ie,~$w \neq
-1$) would provide much-needed experimental input into its nature.
\end{itemize}
\noindent
These supply more than enough motivation for continued efforts to test
the cosmological model and improve its accuracy.  Over the coming
years, there are a wide range of new observations that will bring
further precision to cosmological studies. Indeed, there are far too
many to mention them all here, and so we highlight just a few areas.

The CMB observations will improve in several directions.  A current
frontier is the study of polarization, for which power spectrum
measurements have now been made by several experiments. Detection of
primordial $B$-mode anisotropies is the next major goal and a variety
of projects are targeting this, though theory gives little guidance as
to the likely signal level.  Future CMB projects that are approved
include
\index{CMB!future Simons observatory}%
the Simons Observatory and {\it LiteBIRD}.

An impressive array of cosmology surveys are already operational,
under construction, or proposed, including the ground-based Hyper
Suprime Camera (HSC) and Rubin-LSST imaging surveys, spectroscopic
surveys such as DESI, and space missions {\it Euclid} (succesfully
launched in July 2023) and the \textit{Roman} Space
Telescope.

An exciting area for the future is radio surveys of the redshifted
21-cm line of hydrogen. Because of the intrinsic narrowness of this
line, by tuning the bandpass the emission from narrow redshift slices
of the Universe will be measured to extremely high redshift, probing
the details of the reionization process at redshifts up to perhaps 20,
as well as measuring large-scale features such as the BAOs.  LOFAR and
CHIME are the first instruments able to do this and have begun
operations. In the longer term, the Square Kilometre Array (SKA) will
take these studies to sophisticated levels.

The development of the first precision cosmological model is a major
achievement. However, it is important not to lose sight of the
motivation for developing such a model, which is to understand the
underlying physical processes at work governing the Universe's
evolution. From that perspective, progress has been much less
dramatic. For instance, there are many proposals for the nature of the
dark matter, but no consensus as to which is correct. The nature of
the dark energy remains a mystery. Even the baryon density, now
measured to an accuracy of a percent, lacks an underlying theory able
to predict it within orders of magnitude. Precision cosmology may have
arrived, but at present many key questions remain, to motivate and
challenge the cosmology community.

%\smallskip
%\noindent{\bf References:}
%
%\smallskip
%\parindent=20pt
%\ListReferences
%\vfill\eject

%\endRPPonly

% References
% ----------
% The following line includes your bibliography using BibTeX. In case you do not
% yet use BibTex, you can put your bibliography below (using a series of \bibitem entries).
% Please  note that using BibTeX will become mandatory in the future.
\IfFileExists{hubble.bib}{\putbib[hubble]}{}

\end{bibunit}
\fi

% Generate index for draft mode
\ifdefined\isdraft
\clearpage
\renewcommand{\twocolumn}[1][]{
     \twocolumngrid
     #1
}
\printindex
\fi

\end{document}